\documentclass[11pt,a4paper]{article}  

\usepackage[a4paper, top=1in, left=1in, right=1in, bottom=1in]{geometry}

\date{}
\usepackage{braket}
\usepackage{graphicx}
\usepackage{amsmath,amssymb,amsfonts}
\usepackage{hyperref}
\usepackage{bbold}
\usepackage{caption}
\usepackage{amsmath}  
\usepackage{amssymb}   
\usepackage{braket}
\usepackage{amsthm}  
\usepackage{lipsum}

\usepackage{amssymb}
\usepackage{changepage}
\usepackage{breqn}
\usepackage{amsmath}
\usepackage{amsfonts}
\usepackage[utf8]{inputenc}
\usepackage[T1]{fontenc}
\usepackage{amsmath}
\usepackage{amsfonts}
\usepackage{amssymb}
\usepackage{dsfont}
\usepackage{graphicx}
\usepackage[utf8]{inputenc}
\usepackage{amssymb}

\usepackage{changepage}

\usepackage{changepage}
\usepackage{breqn}
\usepackage{amsmath}
\usepackage{amsfonts}
\usepackage{tikz}
\usepackage[mathscr]{euscript}
\usepackage{yfonts}
\usepackage{tikz-cd}

\usetikzlibrary{positioning}

\usepackage{graphicx}
\usepackage[utf8]{inputenc}
\usepackage{amssymb}
\usepackage{amsthm}  
\usepackage{subcaption} 
\usepackage{url}
\usepackage{tikz}

\usepackage{graphicx}
\usepackage{bm}
\usepackage{latexsym}
\usepackage{amsmath}
\usepackage{bbm}
\usepackage{dsfont}

\usepackage{graphicx}
\usepackage{bm}
\usepackage{latexsym}
\usepackage{amsfonts}
\usepackage{amssymb}
\usepackage{pax}
\usepackage{wasysym}

\usepackage{amsfonts}
\usepackage{amssymb}

\usepackage{pax}
\usepackage{wasysym}

\title{\large New exact solutions for microscale gas flows}

\begin{document}
\maketitle
\date{}
Hollis Williams$^1$

\text{School of Engineering},\text{ University of Warwick},\text{ Coventry} \text{ CV4} \text{ 7AL}, \text{UK}
\newline
\newline
Contact email: $^1$Hollis.Williams@warwick.ac.uk
\newline

	\begin{abstract}\noindent
We present a number of exact solutions to the linearised Grad equations for non-equilibrium rarefied gas flows and heat flows.  The solutions include the flow and pressure fields associated to a point force placed in a rarefied gas flow close to a no-slip boundary and the temperature field for a point heat source placed in a heat flow close to a temperature jump boundary.  We also derive the solution of the unsteady Grad equations in one dimension with a time-dependent point heat source term and the Grad analogue of the rotlet, a well-known singularity of Stokes flow which corresponds to a point torque.

	\end{abstract}

\section{Introduction}
\noindent
Rarefied microscale gas flows are present in a wide variety of applications.  In biomedical science, they are used in the study of the transport and inhalation of small droplets which can potentially carry respiratory viruses when they are expelled from an infected person during coughing or sneezing [1].  They are also applied in the design of microelectromechanical system (MEMS) devices where the relevant gas flows are slow rarefied flows [2].  In astrophysics, there are certain bodies in the Solar System whose atmospheres are of extremely low density such that the gases in question must be considered as rarefied gases [3].  It is widely known that microscale dilute gas flows cannot be accurately described with the traditional fluid dynamics framework based on the Navier-Stokes-Fourier closure and that the Navier-Stokes-Fourier equations can fail to capture the relevant phenomena even in a qualitative manner [4].  The study of rarefied gas motion goes back at least as far as the work of Maxwell, who used kinetic theory to study gas flows induced by temperature gradients [5].  There have since been many studies of rarefied gas flows and in particular, there has been a great deal of interest in thermophoresis on a sphere in a steady-state rarefied gas flow (thermophoresis being the force experienced by solid particles or surfaces in a rarefied gas in the presence of a temperature gradient) [6, 7].  Interest in rarefied gas flows in idealised geometries apart from microflows around isolated spherical particles has recently renewed, with examples including the rarefied gas flow between two infinite parallel plates [8, 9] and the drag force on a sphere moving close to a planar boundary through a highly rarefied gas [10].  Modelling of rarefied gas flows is generally complicated even for steady-state flows in simple geometries because the usual continuum NSF equations lose their accuracy as the Knudsen number $\text{Kn}$ (the ratio of the mean free path in the gas $\lambda$ to the characteristic length scale $L$ for the problem) becomes larger.    

For Knudsen numbers bigger than $0.01$, the NSF equations typically give a picture which is not even qualitatively correct, missing out on the existence of effects such as heat flow from colder regions of the gas to warmer regions [11].  Once the gas flow becomes moderately rarefied, the Boltzmann equation is the only model which gives accurate answers, and must in general be solved numerically.  Numerical solution of the Boltzmann equation can be implemented via direct simulation Monte Carlo (DSMC) [12] or deterministic methods [13], but these are usually prohibitively expensive in computational cost for the flows which interest us.  DSMC becomes computationally expensive when used to model low-speed microflows where the flow velocity is much smaller than the thermal velocity and is particularly expensive for Knudsen numbers in the so-called transition regime between $0.01$ and $1$ at low speeds.  In the flows in MEMS which interest us in applications, the signal to noise ratio is too low to make solution by DSMC computationally viable [13, 14].  Our statements on computational cost depend on the assumption that the interaction between the particles is described by a simple inverse power law potential rather than the more realistic Lennard-Jones potential but this is a standard and widely used assumption in the literature (as is our assumption that the gas is monatomic) [15].  There are also methods which use the smallness of the Knudsen number to greatly increase the efficiency of numerical solution of the Boltzmann equation for moderately small Knudsen numbers [16].

One possibility for effective modelling of rarefied gas flows in the transition regime which also might have some physical insight is to create continuum equations using moment approximations to the Boltzmann equation.  The original approximation of this kind was made by Grad who expanded the distribution function in Hermite polynomials and derived $13$-moment equations known as the Grad or G13 equations [17, 18].  The Grad equations suffer from a number of deficiencies, including inability to capture the effect of Knudsen boundary layers.  Struchtrup and Torrilhon proposed a regularisation of the Grad equations to overcome these issues which are known as the R13 equations [19].  We will not attempt to explain the method of moments here or justify the Grad equations, as there are various detailed reviews in the literature which the reader could consult [20].  Informally, one can think of the Grad equations as an approximation to the Boltzmann equation which is suitable for low-speed microflows at the transitional Knudsen numbers which we are going to study.  Moment equations do not currently have tractable methods of numerical solution in three dimensions.  Techniques do exist for certain two-dimensional flows, but these are difficult to apply to low-speed external flows [21].  However, given that the flows which interest us are slow creeping flows, one can make significant simplifications by considering the linearised forms of the Grad and R13 equations and in this paper we will only work with the linearised versions.  Even in linearised form, solution of the R13 equations is non-trivial and typically involves exact symmetry assumptions [22].  This is the main reason for studying analytic solutions to the Grad equations rather than the R13 equations, and there are already some existing examples in the literature.    

We will now summarise the previous work which has been done in this respect.  In an important study of thermophoresis on spherical particles, Young examined several models for particle thermophoresis and solved the Grad equations with spherical symmetry to obtain an analytic solution for the drag force on a sphere (Young explicitly states that he has chosen to study the Grad equations because of the complexity of the R13 equations even in their linearised form) [23].  Young also showed that the G13 method generates a hierarchy of expressions for the thermophoretic force on a sphere at low Knudsen numbers which includes all the classic results and that the G13 solution can be used to derive an interpolation formula for the transition regime [23].  The relevance of the G13 method is further clarified in [20].  Before Young, Dwyer attempted to give a theory for the thermophoretic force on a spherical particle based on an analytic solution of the Grad equations [24].  Lockerby and Collyer derived Green functions for the G13 equations, where a Green function is taken to be the flow response to a Dirac delta forcing term.  We will discuss these solutions in more detail in Section 2 [14].  These solutions are useful from the numerical point of view because one can exploit linearity of the equations and represent a flow field with a superposition of Green functions [25].  The number of analytic solutions to the Grad equations is still somewhat limited given the physical interest of the equations and that these solutions are mostly restricted to the study of thermophoresis on spherical particles or to solutions corresponding to points placed in an infinite dilute gas.  

The aim of this work is to study the classical analytic solutions and techniques from potential theory for Stokes flow and investigate the extent to which these techniques can be carried over to slow, steady-state rarefied gas flows.  In the process, we hope to derive new exact and integral solutions which can be employed in numerical schemes or used to model steady-state dilute gas flows in certain geometries which are idealised but which at least include some kind of boundary.  Given that the equations of motion and the boundary conditions for the analogous boundary value problems are more complicated, it would be natural to expect that these techniques will not carry over entirely and that it might not be possible to derive the analogue for certain analytic solutions of Stokes flow.  This turns out to be the case, as we will see shortly.  

The structure of the paper is as follows.  In Section 2, we will fix notation and outline the methods, equations, boundary conditions and solutions which we will use in the rest of the paper.  In Sections 3 and 4, we will consider some applications of the method of images.  The basic idea of this method in the context of fluid dynamics is to consider the flow fields generated by a point forcing or a point heat source close to a boundary surface in terms of an 'image system' of singularities on the other side of the wall.  The related fields can then be found by using Fourier analysis to obtain the image system which satisfies the required boundary conditions.  This method has not before been applied to dilute microscale gas flows as far as we know.  Using the method of images we will derive the velocity and temperature fields due to point forcings and point heat sources in the presence of and without velocity slip and temperature jump at the boundary.  The purpose of this is two-fold: to obtain a better physical understanding of velocity (temperature) fields due to a point forcing (point heat source) in dilute microscale gas flows and heat flows close to a plane boundary, and to derive novel exact solutions which can be implemented in the study of such flows in simple geometries and problems.  This could potentially enable us to better understand the movement of slender bodies or small aspherical particulates in rarefied microscale gas flows close to walls.  The solutions could also be employed in certain numerical schemes [14, 25]. 

In Section 5, we derive a novel solution for the unsteady Grad equations given a time-dependent point heat source.  The motivation for this is to try to extend our understanding of dilute microscale gas flows slightly beyond the usual steady-state case and to obtain a solution which complements the existing time-dependent fundamental solutions obtained for unsteady Stokes flow [26].  This solution could be implemented in a numerical method which uses time-dependent fundamental solutions and used to model an unsteady heat flow in a very simple geometry [27].  Our assumption of one dimension seems prohibitive, but is quite typical in the existing literature [28].  Many of the solutions which we derive refer to a flow with constant velocity and pressure.  This is obviously quite a restrictive condition, but does correspond to some physically realistic scenarios of current interest (for example, Fourier flow, where one has a static gas between two stationary parallel plates and the flow structure as it exists is due only to the temperature difference between the plates) [29].  We will refer to these types of flow as heat flows to emphasise that the medium is not moving.

Finally, in Section 6, we derive the analogue of the 'rotlet' from studies of Stokes flow (the rotlet being a point torque rather than a point force).  As well as extending the result from Stokes flows to rarefied microscale gas flows for the sake of interest, this result has potential applications in biology and biomedical science since micro-organisms can exhibit helical beating motions as they move close to flat walls, in which case angular momentum needs to be accounted for [30].  To use the rotlet which we derive in applications, one would need to place it close to a no-slip boundary and determine the associated velocity and pressure fields.  One might wonder if this could be made more realistic by placing the rotlet close to a partial-slip boundary, but we show in Section 4 that obtaining the relevant fields even for the simpler case of a point forcing in a rarefied gas flow close to a partial-slip plane boundary is unfortunately intractable.

\section{Preliminaries}
\noindent
For the sake of clarity, we will now fix all the notation and equations which are used in the rest of the paper so that the reader can refer back to them as necessary.  The basic geometry which we will consider for all the results in Sections 3 and 4 is shown in Figure 1.  The domain on which our PDEs are defined is the upper half-space given by the set of all triples $S = \{ (x,y,z) : z > 0\}$ and the boundary of the domain (denoted in gray in Figure 1) is the flat infinite plane given by all triples $\{ (x,y,z) : z = 0\}$.  At a height $z=h$ above the plane, one then places either a point source of force $\textbf{f} \delta(\textbf{r})$ or a point source of heat $g \delta( \textbf{r})$ at the point $(x,y,z) = (0,0,h)$, where $\textbf{f}$ is the point force vector, $g$ is the strength of the heat source, $\delta$ denotes the Dirac delta function, and $\textbf{r} = (x,y,z-h)$ is the vector from the source to the point of observation.  On the opposite side of the wall outside of the domain on which the PDEs are defined, one imagines that there is a mirror image point source of force $\overline{\textbf{f}}  \delta( \overline{\textbf{r}})$ or a point source of heat $g \delta (\overline{\textbf{r}})$, where $\overline{\textbf{f}}$ is the associated point force vector chosen to satisfy the no-flux boundary condition, and $\overline{\textbf{r}} = (x,y,z+h)$ is the position vector for the point at which the image source is observed.  The idea of the method of images is to decompose the velocity field $\textbf{v}$ into three parts

\[  \textbf{v} = \textbf{u} + \textbf{u}' + \textbf{u}'',  \tag{1} \]

\begin{figure*}
\centering
\includegraphics[width=0.7\textwidth]{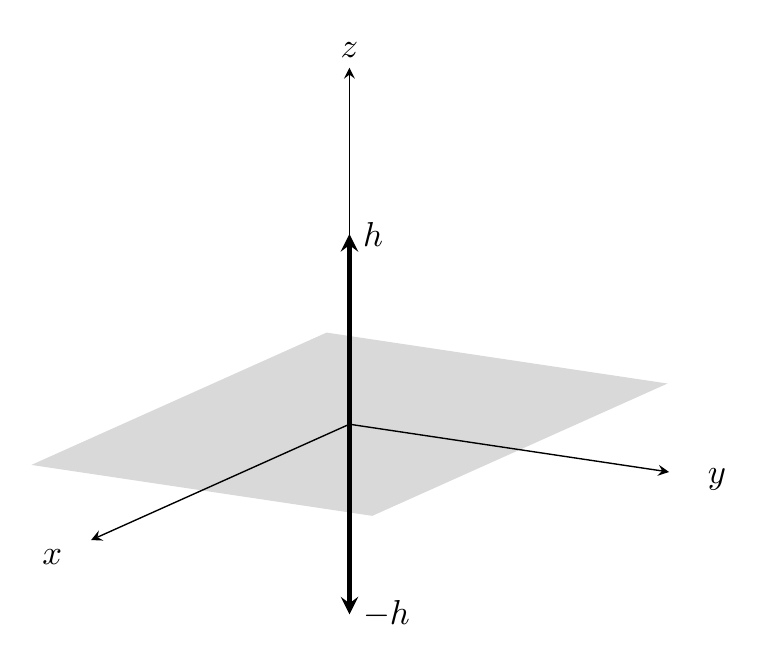} 
\caption{The point force or heat source is located at a height $h$ above the boundary at $z=0$ and there is a mirror image point force or heat source whose strength is equal in magnitude at a height $h$ below the boundary outside the domain in which the PDEs are defined.  Note that the point source of force can be perpendicular to the plane at $z=0$, or parallel to it.}
\end{figure*}

\noindent
where $\textbf{u}$ is the velocity field due to the point forcing placed in the flow, $\textbf{u}'$ is the velocity field due to the mirror image point forcing placed on the other side of the wall, and $\textbf{u}''$ is the unknown velocity field due to some system of singularities on the other side of the wall.  The basic reason for needing more singularities in the image system apart from the mirror image point force is that velocity is a vector quantity and not a scalar, hence an image point force by itself would not be able to cancel those components of the flow which are both tangential and normal to the boundary surface.  Note that the prime does not denote differentiation and we hope that this is not misleading, as we feel that it is more confusing to denote the different parts of the total velocity field $\textbf{v}$ with different letters.  $\textbf{v}$ as a whole is required to solve the boundary value problem, so one uses Fourier analysis to find the unknown velocity field $\textbf{u}''$ and finally adds the three parts together to obtain the total flow field due to a point force or heat source close to the wall.  The analysis in question requires two-dimensional Fourier transforms.  For reference, the Fourier transform and the inverse Fourier transform are defined via

\[ \hat{{u}}_i ( k_1, k_2, z) = \frac{1}{2 \pi} \int \int   u_i (x,y,z) e^{i k_1 x + i k_2 y} \: \text{d}x \: \text{d}y, \tag{2a}\]

\[ u_i ( x, y, z) = \frac{1}{2 \pi} \int \int  \hat{{u}}_i ( k_1, k_2, z) e^{-i k_1 x - i k_2 y} \: \text{d}k_1 \: \text{d}k_2.  \tag{2b}\]

\noindent
Useful tables of Fourier transforms can be found at [31, 32].  Quantities with hats indicate Fourier transformed quantities.  In Figure 1, we have mentioned that one must consider separately the cases where the point forcing is perpendicular or parallel to the wall: this is because the presence of the wall breaks isotropy of particles moving close to it.  A superscript $\perp$ for a quantity indicates that we are considering the case where the point forcing is perpendicular to the boundary (in the $z$-direction), and a superscript $\parallel$ for a quantity indicates that we are considering the case where the point forcing is parallel to the boundary (in the $x$-direction).  To be clear, this has nothing to do with the actual components of the velocity field itself.  By the no-flux boundary condition, the $z$-component of the velocity field vanishes at the boundary whether the point forcing is in the $x$- or the $z$-direction.  

\subsection{Stokes Equations with Point Forcing}

We now outline the PDEs, boundary conditions and solutions which we will be referring to throughout the paper.  To begin with, the Stokes equations are given by  

\[ \nabla \cdot \textbf{v} = 0 , \tag{3a}\]

\[ \nabla p + \nabla \cdot \textbf{S} = 0 , \tag{3b}\]

\[ \textbf{S} = - 2 \text{Kn}' \overline{\nabla \textbf{v}}  , \tag{3c}\]

\noindent
where $\text{Kn}'$ is a re-scaled version of the Knudsen number which is relevant for the gas flows which we are studying

\[ \text{Kn}' = \sqrt{\frac{2}{\pi}} \text{Kn}   \tag{4}\]

\noindent
and the overline denotes the symmetric trace-free part of a tensor.  The equations which we consider are dimensionless and linearised about an equilibrium state which is defined by a constant reference density $\rho_0$ and a reference temperature $\theta_0$.  We only consider small deviations away from equilibrium in this work.  The relations between the variables for dimensionless deviation away from the equilibrium state (denoted with tilde symbols) and variables with dimension are

\[ \theta = \theta_0 (1 + \tilde{\theta}), \:\:\:\: \:\:\:\: \rho = \rho_0 ( 1 + \tilde{\rho}), \:\:\:\: \:\:\:\: p = p_0 (1 + \tilde{p} ), \:\:\:\: \:\:\:\: \textbf{r} = L \tilde{\textbf{r}}, \tag{5a}\]

\[ \textbf{v} = \sqrt{R \theta_0} \tilde{\textbf{v}},  \:\:\:\: \:\:\:\: \textbf{q} = \rho_0 (R \theta_0)^{3/2} \tilde{\textbf{q}} ,\:\:\:\: \:\:\:\:  \textbf{S} = \rho_0 R \theta_0 \tilde{\textbf{S}}, \tag{5b} \]

\noindent
where $L$ is a characteristic length scale, $p_0$ is the equilibrium pressure and $R$ is the ideal gas constant.  In the rest of this work, tilde symbols will not be shown and all variables will be dimensionless unless stated otherwise.

The Green function for the Stokes equations with a point forcing term on the right-hand side of the momentum equation is extremely well-studied in the literature (recall that the point forcing for the Stokes equations is called the Stokeslet) [33].  The reader should bear in mind that with a point forcing term applied, the momentum equation $(..)$ is actually

\[   \nabla p + \nabla \cdot \textbf{S} = 0  = \textbf{f} \delta( \textbf{r} ).  \tag{6}\]

\noindent
Assuming that the pressure and the magnitude of the velocity and stress tensors vanish as one goes to infinity, it is well-known that one obtains [32, 14]

\[\textbf{v} = \textbf{f} \cdot \textbf{G}, \:\: \: \: \:\: \: \:  p = \frac{\textbf{f} \cdot \textbf{r}}{4 \pi | \textbf{r}|^3} , \:\: \: \: \:\: \: \:  \textbf{S} = - \frac{\textbf{f} \cdot \textbf{r}}{4 \pi} \Bigg( \frac{\textbf{I}}{|\textbf{r}|^3} - \frac{3 \textbf{r} \textbf{r}}{| \textbf{r} |^5} \Bigg) \tag{7}\]

\noindent
where $\textbf{I}$ is the identity matrix and the Green function $\textbf{G}$ is given by

\[ \textbf{G} = \frac{1}{8 \pi \text{Kn}'} \Bigg( \frac{\textbf{I}}{|\textbf{r}|} + \frac{\textbf{r} \textbf{r}}{|\textbf{r}|^3} \Bigg) . \tag{8}\]

\noindent
Note that this is the free-space solution and that we have not yet introduced a boundary condition at $z=0$.  Using the parallel and perpendicular notation which we mentioned above, one has for parallel and perpendicular Stokeslets

\[  \textbf{v}^{\perp} = f \textbf{G}_{x} , \tag{9a}\]

\[ \textbf{v}^{\parallel} = f \textbf{G}_{z}  ,\tag{9b}\]

\noindent
where $f$ is the magnitude of the force, so one takes the component of the Green function which corresponds to the direction of the point forcing.  For the velocity field due to the mirror image point forcing, the sign in front of the force is reversed if one takes the force in the perpendicular direction

\[  \textbf{v}'{^{\perp}} = -f \textbf{G}_{x} ,\tag{10a}\]

\[ \textbf{v}'{^{\parallel}} = f \textbf{G}_{z}  ,\tag{10b}\]

\noindent
which often leads to some cancellations in the computations.  The sign of the force is chosen in this way to enforce the no-flux condition at the boundary.  The no-slip boundary condition is simply the requirement that every component of the velocity vanish at the boundary

\[ \textbf{v} = 0 \tag{11} \]

\noindent
and for a stationary wall the slip boundary condition in terms of the stress tensor $\textbf{S}$ is [14]

\[ \textbf{v} = - \sqrt{\frac{\pi}{2}} \textbf{n} \cdot \textbf{S} \cdot (\textbf{I} - \textbf{n} \textbf{n} ). \tag{12} \]

\noindent
where $\textbf{n}$ is the unit normal vector to the boundary surface.  In our case, the boundary surface is a flat plane at $z=0$ and the unit normal vector is the unit vector in the $z$-direction, so the slip boundary condition in terms of each component of the velocity becomes

\[ v_1 = - \sqrt{\frac{\pi}{2}} S_{31} , \tag{13a}\]

\[ v_2 = - \sqrt{\frac{\pi}{2}} S_{32} , \tag{13b}\]

\[ v_3 = 0 . \tag{13c}\]

\subsection{Stokes Equations with Point Heat Source}

One can instead look for solutions when the conservation equations are subject to a steady-state point heat source rather than a point force source (in [14], this point heat source is called a thermal Stokeslet).  In terms of the stress tensor and heat flux, the conservation equations are

\[  \nabla \cdot \textbf{v} = 0, \tag{14a}\]

\[ \nabla p + \nabla \cdot \textbf{S} = 0 , \tag{14b}\]

\[\nabla \cdot \textbf{q} = g \delta ( \textbf{r}) , \tag{14c}\]

\noindent
where $\textbf{q}$ is the heat flux and $g$ is the strength of the point heat source.  The equations for the stress tensor and heat flux are given by the NSF closure:

\[ \textbf{S} = - 2 \text{Kn} ' \overline{\nabla \textbf{v}}  , \tag{15a}\]

\[ \textbf{q} = - \frac{15}{4} \text{Kn}' \nabla \theta. \tag{15b}\]

\noindent
In the literature, one then typically looks for isobaric zero flow solutions such that $\textbf{v} = p = 0$ which gives us the equations

\[\nabla \cdot \textbf{q} = g \delta ( \textbf{r}) , \tag{16a}\]

\[ \textbf{q} = - \frac{15}{4} \text{Kn}' \nabla \theta. \tag{16b}\]

\noindent
Taking the boundary condition to be that the temperature $\theta$ scaled by the mass goes to its equilibrium value at infinity, one then finds that for the thermal Stokeslet applied to an infinite rarefied gas, the temperature $\theta$ and the heat flux $\textbf{q}$ are

\[ \theta = \frac{g}{15 \text{Kn}' \pi |\textbf{r}|}, \tag{17a}\]

\[\textbf{q} =  \frac{g \textbf{r}}{4 \pi|\textbf{r}|^3}  . \tag{17b} \]

\noindent
The no temperature jump boundary condition is the requirement that the temperature vanish at the boundary

\[ \theta = 0 \tag{18} \]

\noindent
and the temperature jump boundary condition is 

\[ \theta = - \frac{1}{2} \sqrt{\frac{\pi}{2}} \textbf{n} \cdot \textbf{q}  . \tag{19}\]

\noindent
Since the unit normal vector is the unit vector in the $z$-direction, the temperature jump boundary condition becomes

\[\theta = - \frac{1}{2} \sqrt{\frac{\pi}{2}} q_3  . \tag{20}\]

\subsection{Grad Equations with Point Forcing}

In this work, we are primarily interested in the Grad equations for rarefied gas flows.  For a point forcing applied to the momentum equation, these consist of the usual equations for conservation of mass, momentum, and internal energy

\[  \nabla \cdot \textbf{v} = 0 , \tag{21a}\]

\[ \nabla p + \nabla \cdot \textbf{S} = \textbf{f} \delta (\textbf{r}) , \tag{21b}\]

\[ \nabla \cdot \textbf{q} = 0, \tag{21c}\]

\noindent
combined with equations for the stress tensor and the heat flux given by Grad's closure

\[ \textbf{S} = - 2 \text{Kn}' \overline{\nabla \textbf{v}} - \frac{4}{5} \text{Kn}' \overline{\nabla{\textbf{q}}} , \tag{22a}\]

\[ \textbf{q} = - \frac{15}{4} \text{Kn}' \nabla \theta - \frac{3}{2} \text{Kn}' \nabla \cdot \textbf{S} . \tag{22b}\]

\noindent
The point forcing in the case of the Grad equations is known as the Gradlet [14].  Solving these equations for an infinite rarefied gas, one finds that the velocity and pressure fields associated to a Gradlet in an infinite domain far away from boundaries are

\[  \textbf{v} = \frac{\textbf{f}}{8 \pi \text{Kn}'} \Bigg( \frac{\textbf{I}}{| \textbf{r}|} + \frac{ \textbf{r} \textbf{r}}{|\textbf{r}|^3} \Bigg)  - \frac{3 \text{Kn}' \textbf{f}}{20 \pi} \Bigg( \frac{\textbf{I}}{| \textbf{r}|^3} - \frac{3 \textbf{r} \textbf{r}}{|\textbf{r}|^5} \Bigg),\tag{23a}\]

\[ p = \frac{\textbf{f} \cdot \textbf{r}}{4 \pi |\textbf{r}|^3}  .\tag{23b}\]

\noindent
The slip boundary condition for the Grad equations is more complicated than the slip boundary condition for the Stokes equations and includes a term which depends on the heat flux [34]:

\[ \textbf{v}  = - \sqrt{\frac{\pi}{2}} \textbf{n} \cdot \textbf{S} \cdot (\textbf{I} - \textbf{n} \textbf{n})  - \frac{1}{5} \textbf{q} \cdot (\textbf{I} - \textbf{n} \textbf{n}). \tag{24} \]

\noindent
For a flat plane as the boundary, these become

\[v_1 =  - \sqrt{\frac{\pi}{2}}S_{3 1} - \frac{1}{5}q_{1}  , \tag{25a}\]

\[v_2 = - \sqrt{\frac{\pi}{2}}S_{3 2} - \frac{1}{5}q_{2}  , \tag{25b}\]

\[v_3 = 0 . \tag{25c} \]

\subsection{Grad Equations with Point Heat Source}

As with the Stokes equations, one can instead consider the Grad equations with a point heat source applied to the conservation equations:

\[  \nabla \cdot \textbf{v} = 0 , \tag{26a}\]

\[ \nabla p + \nabla \cdot \textbf{S} = 0, \tag{26b}\]

\[ \nabla \cdot \textbf{q} = g \delta (\textbf{r}), \tag{26c}\]

\[ \textbf{S} = - 2 \text{Kn}' \overline{\nabla \textbf{v}} - \frac{4}{5} \text{Kn}' \overline{\nabla{\textbf{q}}} , \tag{26d}\]

\[ \textbf{q} = - \frac{15}{4} \text{Kn}' \nabla \theta - \frac{3}{2} \text{Kn}' \nabla \cdot \textbf{S} . \tag{26e}\]

\noindent
In this case, the point heat source is known as a thermal Gradlet.  As before, it is usual to simplify by taking $\textbf{v} =p =0$ which gives us

\[ \nabla \cdot \textbf{S} = 0, \tag{27a}\]

\[\nabla \cdot \textbf{q} = g \delta(\textbf{r}), \tag{27b} \]

\[ \textbf{S} = -\frac{4}{5} \text{Kn}' \overline{\nabla \textbf{q}}, \tag{27c}\]

\[ \textbf{q} = - \frac{15}{4} \text{Kn}' \nabla \theta . \tag{27d}\]

\noindent
One then can solve to find the heat flux $\textbf{q}$, the stress tensor $\textbf{S}$ and the temperature $\theta$ rescaled by the mass for a point heat source in an infinite rarefied stationary gas [14]:

\[ \theta = \frac{g}{15 \text{Kn}' \pi |\textbf{r}|}, \tag{28a}\]

\[\textbf{q} =  \frac{g \textbf{r}}{4 \pi|\textbf{r}|^3}  , \tag{28b}\]

\[ \textbf{S} = - \frac{\text{Kn}' g}{5 \pi} \Bigg( \frac{\textbf{I}}{| \textbf{r}|^3} - \frac{3 \textbf{r} \textbf{r}}{| \textbf{r}|^5}  \Bigg) . \tag{28c}\]

\noindent
Again, as before, there is some additional complexity in the temperature jump boundary condition for the Grad equations which now involves a term which depends on the stress tensor [34]:  

\[ \theta = - \frac{1}{2}  \sqrt{\frac{\pi}{2}} \textbf{n} \cdot \textbf{q} - \frac{1}{4} \textbf{n} \cdot \textbf{S} \cdot \textbf{n}  . \tag{29}\]

\noindent
For a flat surface, this simplifies to

\[\theta = -\frac{1}{2} \sqrt{\frac{\pi}{2}} q_3 - \frac{1}{4} S_{33}  . \tag{30}\]
 
\noindent
The coupling between the stress tensor and the heat flux is an effect which is present in rarefied gas flows but not in standard Stokes flows.

\section{Exact Solutions from the Method of Images}

\noindent
The basic question is to what extent the fields due to a point force or a point heat source placed in an infinite rarefied gas are modified in the presence of a flat planar boundary, and this is what we will now proceed to study in some detail via the method of images.  Along the way, we will also derive some solutions for Stokes flows, some of which are new as far as we are aware. 

\subsection{Solutions for No-slip and No Temperature Jump Boundaries}
\noindent
We will begin by focussing on the no-slip and no temperature jump boundary conditions given by (11) and (18).  The situation is simplest for a boundary value problem where the temperature is prescribed to vanish at the boundary.  In this setting, the image system required to solve the boundary value problem is just the mirror image of whatever was placed in the domain on the other side of the wall.  

As an example, we start with the temperature and heat flux fields associated to a point heat source placed in isobaric stationary-state Stokes flow close to a no temperature jump surface (equations (14), solutions (17) and boundary condition (18)).  As emphasised before, this type of flow would be better characterised as a heat flow, since the medium is not moving.  We will abuse notation slightly and decompose the total temperature field as $\theta + \theta' + \theta''$, where $\theta$ denotes the temperature field for the point heat source placed in the domain close to the boundary, and does not denote the total temperature field unless explicitly stated otherwise (this should only be at the end of a section when we are ready to write down total fields).  Hopefully this is not too confusing, as it saves on having to introduce extra letters in the notation.  If the total temperature field satisfies the no temperature jump boundary condition, then we must have

\[ \theta + \theta' + \theta'' = 0 .\tag{31}\]

\noindent
After taking Fourier transforms of each quantity, we have 

\[ \hat{\theta} +\hat{\theta}' + \hat{\theta}'' = 0. \tag{32}\]

\noindent
$\theta$ is given by (17a), so taking the Fourier transform we have

\[ \hat{\theta} =  \frac{g}{15 \text{Kn}' \pi} \frac{1}{k} e^{-kz} , \tag{33}\]

\noindent
where $k = \sqrt{k_1^2 + k_2^2}$.  The next point also has some potential for confusion, so we will clarify.  The unknown temperature field $\theta''$ and unknown heat flux $\textbf{q}''$ by themselves satisfy equations (14), but without the point heat source term on the right-hand side of the conservation equation:

\[\nabla \cdot \textbf{q} = 0 , \tag{34a}\]

\[ \textbf{q} = - \frac{15}{4} \text{Kn}' \nabla \theta. \tag{34b}\]

\noindent
This needs to be made clear as it may be used implicitly in the rest of the text.  It should be clear from a physical point of view because the unknown fields are associated with a fictitious system of singularities on the other side of the wall outside the physical domain.  To find $\hat{\theta}''$, one takes two-dimensional Fourier transforms of the above equations and solves the resulting differential equations.  The above equations simplify to

\[ \Delta \theta'' = 0 . \tag{35}\]

\noindent
The two-dimensional Fourier transform of this is 

\[ \frac{\partial^2 \hat{\theta}''}{\partial z^2} - k^2 \theta''=0  , \tag{36}\]

\noindent
where $k^2 = k_1^2 + k_2^2$.  The solution to this differential equation is

\[ \hat{\theta}'' = A e^{-kz}  , \tag{37}\]

\noindent
where $A$ is a constant to be determined and the positive exponential term has been neglected because the temperature goes to zero at spatial infinity.  Substituting all the hatted quantities into the Fourier transformed boundary condition at $z=0$, we have

\[A = - \frac{2g}{15 \text{Kn}' \pi} \frac{1}{k} \tag{38} \]

\noindent
so

\[ \hat{\theta} '' = - \frac{2g}{15 \text{Kn}' \pi} \frac{1}{k} e^{-kz} . \tag{39}\]

\noindent
From equation (14c), we have 

\[ q_1'' = - \frac{15}{4} \text{Kn}' \frac{\partial \theta}{\partial x} ,  \: \: \: \: q_2'' = - \frac{15}{4} \text{Kn}' \frac{\partial \theta}{\partial y} , \: \: \: \: q_3'' = - \frac{15}{4} \text{Kn}' \frac{\partial \theta}{\partial z}  \tag{40} \]  

\noindent
Taking Fourier transforms as usual, we have

\[ \hat{q}''_1 = \frac{15 i k_1 }{4} \text{Kn}' \hat{\theta }  = - \frac{g}{2 \pi} \frac{i k_1}{k} e^{-kz}, \tag{41a}\]

\[ \hat{q}''_2= \frac{15 i k_2 }{4} \text{Kn}' \hat{\theta }  = - \frac{g}{2 \pi} \frac{i k_2}{k} e^{-kz}, \tag{41b}\]

\[ \hat{q}''_3 = \frac{15}{4}k \text{Kn}' A e^{-kz} = - \frac{g}{2 \pi} e^{-kz} . \tag{41c}\]

\noindent
We finish by taking the inverse Fourier transforms to arrive at

\[ \theta'' = - \frac{2g}{15 \text{Kn}' \pi} \frac{1}{|\textbf{r}|}  , \tag{42a}\]

\[ \textbf{q}'' =-\frac{g \textbf{r}}{2 \pi} \frac{1}{|\textbf{r}|^3},  \tag{42b} \]

\noindent
and adding the unknown fields to the fields due to the point heat source in the domain and the mirror image heat source on the other side to obtain the expected total fields for a heat source close to a no temperature jump surface

\[\theta = \frac{g}{15 \text{Kn}' \pi} \Bigg(   \frac{1}{|\overline{\textbf{r}}|} - \frac{1}{|\textbf{r}|}  \Bigg),  \tag|, \tag{43a}\]

\[\textbf{q} = \frac{g \textbf{r}}{4 \pi} \Bigg(   \frac{1}{|\overline{\textbf{r}}|^3} - \frac{1}{|\textbf{r}|^3}  \Bigg). \tag|. \tag{43b}\]

\noindent
We have been somewhat explicit here, but hope that the method of calculation should be clear in the rest of the paper even if some steps are missed out for brevity.  Similar computations confirm that the fields due to a point heat source placed in a isobaric stationary-state rarefied gas flow close to a no temperature jump surface (equations (27), solutions (28) and boundary condition (18)) are equivalent to the ones obtained in (43a) and (43b).

\subsection{Velocity Distribution for a Gradlet close to a No-slip Surface}

\noindent
The next problem is to determine the fields due to a point forcing (known in this context as a Gradlet) placed in a rarefied gas flow close to a no-slip boundary (equations (22), solutions (23) and boundary conditions (24)).  As mentioned earlier, we do have to consider as separate cases when the point forcing is perpendicular to the wall or parallel to it, because the wall breaks isotropy of nearby particle motions.  The Grad equations are generally combined with a slip boundary condition.  The reason for this is that gas flows in microscale or nanoscale devices whose dimensions are of the order of the mean free path of the gas molecules typically exhibit a significant amount of gas slip.  However, it is possible to have certain rarefied gas flows where the effect of gas slip can be approximately neglected [35, 36] and in this case, the exact solution which we are about to derive would be very useful and applicable in mesh-free numerical methods [37].

Starting with equations (21) and removing the point forcing term applied to the momentum equation, it can be shown with some substitutions [23] that

\[ \nabla^2 p'' =0, \tag|{44a}\]

\[ \nabla p'' = \text{Kn}' \nabla^2 \textbf{u}'', \tag|{44b}\]

\[\nabla^2 \textbf{q}''=0 . \tag|{44c}\]

\noindent
Taking Fourier transforms and solving the resulting differential equations, we get

\[ \hat{u}_1'' = \Bigg( \frac{h}{8 \pi \text{Kn}'} \Bigg) (B_1 + i k_1 A z ) e^{-kz}, \tag|{45a}\]

\[ \hat{u}_2'' = \Bigg( \frac{h}{8 \pi \text{Kn}'} \Bigg) (B_2 + i k_2 A z ) e^{-kz}, \tag|{45b}\]

\[ \hat{u}_3'' = \Bigg( \frac{h}{8 \pi \text{Kn}'} \Bigg) (B_3 +  k A z ) e^{-kz}, \tag|{45c}\]

\[ \hat{p} = \Bigg(\frac{h}{4 \pi} \Bigg) A e^{-kz} . \tag|{45d}\]

\noindent
where $h$ is the height at which the Gradlet has been placed above the boundary.  The goal is now to use the boundary conditions and the continuity equation (21a) to determine the constants $A$ and $B_i$.  We will start with the case where the Gradlet is perpendicular to the surface.  The no-slip boundary condition is 

\[ \textbf{u}^{\perp} + \textbf{u}'{^{\perp}} + \textbf{u}''{^{\perp}} = 0 . \tag{46}\]

\noindent
The meaning of the primes in the decomposition of the total velocity field $\textbf{v}$ into three parts is the same as in the previous section and was also explained in the preliminary section.  Next, add the velocity fields (23) for the Gradlet and the mirror image Gradlet when the Gradlet is in the perpendicular direction (ie. one takes the $z$-component of the force).  If one took the point forcing to be in the parallel direction, the expressions would be slightly different as one would take the $x$-component of the force.  This leaves us with

\[  u_1^{\perp} + u_1'{^{\perp}} = - \frac{1}{4 \pi \text{Kn}'} \frac{xh}{r_h^3} - \frac{9 \text{Kn}'}{10 \pi} \frac{x h}{r_h^5} , \tag{47a}\]

\[  u_2^{\perp} + u_2'{^{\perp}} = - \frac{1}{4 \pi \text{Kn}'} \frac{yh}{r_h^3} - \frac{9 \text{Kn}'}{10 \pi} \frac{y h}{r_h^5} , \tag{47b}\]

\[  u_3^{\perp} = 0 ,  \tag{47c}\]

\noindent
where $r_h^2 = x^2 + y^2 + h^2$ and $h$ is always the height at which the point force or point heat source is placed above the boundary.  We Fourier transform this to get

\[\hat{u}_1{^{\perp}} + \hat{u}'_1{^{\perp}} = - i \frac{h}{4 \pi \text{Kn}'} \frac{k_1}{k} e^{-kh} - i \frac{3 \text{Kn}'}{10 \pi} k_1 e^{-kh} , \tag{48a}\]

\[\hat{u}_2{^{\perp}} + \hat{u}'_{2^{\perp}} = - i \frac{h}{4 \pi \text{Kn}'} \frac{k_2}{k} e^{-kh} - i \frac{3 \text{Kn}'}{10 \pi} k_2 e^{-kh} ,\tag{48b}\]

\[\hat{u}_3^{\perp} + \hat{u}'_3{^{\perp}} = 0 .\tag{48c}\]

\noindent
Finally, we add all the hatted quantities at the boundary at $z=0$ to obtain expressions for the coefficients $B_i$:

\[\frac{h}{8 \pi \text{Kn}'} B^{\perp}_1 = -i \frac{h}{4 \pi \text{Kn}'} \frac{k_1}{k} - i \frac{3 \text{Kn}'}{10 \pi} k_1   , \tag{49a}\]

\[ \frac{h}{8 \pi \text{Kn}'} B^{\perp}_2 = -i \frac{h}{4 \pi \text{Kn}'} \frac{k_2}{k}  - i \frac{3 \text{Kn}'}{10 \pi} k_2  \tag{49b} \]

\[ \frac{h}{8 \pi \text{Kn}'} B^{\perp}_3 =0 . \tag|{49c}\]

\noindent
Equation (49c) implies that $B_3^{\perp} =0$, so we only need a third equation to close the system and determine $A^{\perp}$.  This is provided by the continuity equation (21a).  Fourier transforming the continuity equation and making some cancellations, we end up with

\[  A^{\perp} = \frac{i}{k}(k_1 B^{\perp}_1 + k_2 B^{\perp}_2 ). \tag|{50} \]

\noindent
Substituting these constants back into equations (49) and taking inverse Fourier transforms, we obtain the unknown pressure and velocity fields which can be added to the fields for the Gradlet and the mirror image Gradlet to obtain the total fields associated with a point forcing placed in a rarefied gas flow perpendicular to the boundary.

\[  u_1^{{\perp}''} = -\frac{h}{4 \pi \text{Kn}'} \Bigg( \frac{\partial}{\partial z} \Bigg( \frac{h  x}{|\textbf{r}|^3} \Bigg) -  \frac{\partial}{\partial z} \Bigg( \frac{ z x}{|\textbf{r}|^3} \Bigg) \Bigg)  + \frac{3 \text{Kn}'}{10 \pi} \Bigg(  \frac{\partial}{\partial z} \Bigg( \frac{3 z x}{|\textbf{r}|^5} \Bigg) - \frac{\partial }{\partial z} \Bigg( \frac{ 3 h x}{|\textbf{r}|^5} \Bigg) \Bigg) , \tag|{51a}\]

\[ u_2{^{\perp}}'' = -\frac{h}{4 \pi \text{Kn}'} \Bigg( \frac{\partial}{\partial z} \Bigg( \frac{h  y}{|\textbf{r}|^3} \Bigg) -  \frac{\partial}{\partial z} \Bigg( \frac{ z y}{|\textbf{r}|^3} \Bigg) \Bigg)  + \frac{3 \text{Kn}'}{10 \pi} \Bigg(  \frac{\partial}{\partial z} \Bigg( \frac{3 z y}{|\textbf{r}|^5} \Bigg) - \frac{\partial }{\partial z} \Bigg( \frac{ 3 h y}{|\textbf{r}|^5} \Bigg) \Bigg) , \tag|{51b}\]

\[u_3{^{\perp}}'' = -\frac{h}{4 \pi \text{Kn}'} \Bigg(  \frac{\partial}{\partial z}  \Bigg( \frac{hz}{|\textbf{r}|^3} - \frac{1}{|\textbf{r}|} - \frac{z^2}{|\textbf{r}|^3}  \Bigg) \Bigg)+ \frac{3  \text{Kn}'}{10 \pi}\Bigg( \frac{\partial}{\partial z} \Bigg( \frac{3z^2}{|\textbf{r}|^5} - \frac{1}{|\textbf{r}|^3} - \frac{3 h z}{|\textbf{r}|^5}  \Bigg) \Bigg) , \tag|{51c}\]

\[p{^{\perp}}'' = -\frac{h}{4 \pi} \Bigg( \frac{\partial}{\partial z} \Bigg( \frac{2 z}{|\textbf{r}|^3} \Bigg) + \frac{12 \text{Kn}'{^2}}{5 h} \frac{\partial}{\partial z} \Bigg(  \frac{3 z^2}{|\textbf{r}|^5} - \frac{1}{|\textbf{r}|^3} \Bigg)   \Bigg) . \tag|{51d}\]

\noindent
It is straightforward to repeat the analysis for a point forcing in the parallel direction, as this just changes some of the expressions slightly due to certain expressions no longer cancelling out.  The corresponding unknown velocity and pressure fields in this case are

\[ u_1{^{\parallel}}'' = \frac{h}{4 \pi \text{Kn}'} \Bigg( \frac{\partial}{\partial x} \Bigg( \frac{h  x}{|\textbf{r}|^3} \Bigg) -  \frac{\partial}{\partial x} \Bigg( \frac{ z x}{|\textbf{r}|^3} \Bigg) \Bigg)  - \frac{3 \text{Kn}'}{10 \pi} \Bigg(  \frac{\partial}{\partial x} \Bigg( \frac{3 z x}{|\textbf{r}|^5} \Bigg) - \frac{\partial }{\partial x} \Bigg( \frac{ 3 h x}{|\textbf{r}|^5} \Bigg) \Bigg), \tag|{52a}\]

\[ u_2{^{\parallel}}'' =  \frac{h}{4 \pi \text{Kn}'} \Bigg( \frac{\partial}{\partial x} \Bigg( \frac{h  y}{|\textbf{r}|^3} \Bigg) -  \frac{\partial}{\partial x} \Bigg( \frac{ z y}{|\textbf{r}|^3} \Bigg) \Bigg)  - \frac{3 \text{Kn}'}{10 \pi} \Bigg(  \frac{\partial}{\partial x} \Bigg( \frac{3 z y}{|\textbf{r}|^5} \Bigg) - \frac{\partial }{\partial x} \Bigg( \frac{ 3 h y}{|\textbf{r}|^5} \Bigg) \Bigg) , \tag|{52b}\]

\[u_3{^{\parallel}}'' = \frac{h}{4 \pi \text{Kn}'} \Bigg(  \frac{\partial}{\partial x}  \Bigg( \frac{hz}{|\textbf{r}|^3} - \frac{1}{|\textbf{r}|} - \frac{z^2}{|\textbf{r}|^3}  \Bigg) \Bigg) - \frac{3  \text{Kn}'}{10 \pi}\Bigg( \frac{\partial}{\partial x} \Bigg( \frac{3z^2}{|\textbf{r}|^5} - \frac{1}{|\textbf{r}|^3} - \frac{3 h z}{|\textbf{r}|^5}  \Bigg) \Bigg) , \tag|{52c}\]

\[p{^{\parallel}}'' = \frac{h}{4 \pi} \Bigg( \frac{\partial}{\partial x} \Bigg( \frac{2 z}{|\textbf{r}|^3} \Bigg) + \frac{12 \text{Kn}'{^2}}{5 h} \frac{\partial}{\partial x} \Bigg(  \frac{3 z^2}{|\textbf{r}|^5} - \frac{1}{|\textbf{r}|^3} \Bigg)   \Bigg)   .\tag|{52d}\]

\noindent
For the reader who is familiar with the literature on singularities of Stokes flows, what we have shown here is that the image system which is required to solve the boundary value problem is the analogue of the corresponding image system needed to solve the same boundary value problem when one instead has a Stokeslet close to a flat no-slip wall [38].  Recall that the singular point forcing we have defined is only the simplest type of singularity of a flow, and that one can take derivatives to obtain higher-order singularities [39].  In the terminology of the literature on singularities of viscous flows, if the force is in the $z$-direction the additional singularities in the image system on the other side of the wall apart from the the mirror image Gradlet are just the analogous source dipole in the $z$-direction and a $z$-dipole of $z$-Gradlets [30].    

\subsection{Pressure Distribution for a Stokeslet perpendicular to a Slip Surface}

\noindent
We will now derive the total pressure field due to a point forcing in a Stokes flow which is close to a slip boundary and perpendicular to the boundary (equations (3), solutions (7) and boundary conditions (12)).  Taking Fourier transforms of equation (5c) for the unknown stress tensor field, we obtain

\[ \hat{S}''_{1 3}{^{\perp}} =  \hat{S}''_{3 1}{^{\perp}} = -\text{Kn}' \Bigg( \frac{\partial \hat{u}_1{^{\perp}}}{\partial z} - i k_1 \hat{u}_3{^{\perp}} \Bigg) , \tag|{53a}\]

\[ \hat{S}''_{2 3}{^{\perp}} =  \hat{S}''_{3 2}{^{\perp}} = -\text{Kn}' \Bigg( \frac{\partial \hat{u}_2{^{\perp}}}{\partial z} - i k_2 \hat{u}_3{^{\perp}} \Bigg) .\tag|{53b} \]

\noindent
We have only written down the transforms of the components which we will need for our computations.  Making some re-arrangements with the Stokes equations [38], we see that the unknown pressure satisfies the Laplace equation

\[ \Delta p'' =0 , \tag|{54}\]

\noindent
so as before, after taking Fourier transforms and solving the differential equation one obtains

\[ \hat{p}'' = A e^{-kz}.\tag|{55} \]

\noindent
Again, as with Section 3.2, the Fourier transformed components of the unknown velocity field are

\[ \hat{u}_1'' = \Bigg( \frac{h}{8 \pi \text{Kn}'} \Bigg) (B_1 + i k_1 A z ) e^{-kz}, \tag|{56a}\]

\[ \hat{u}_2'' = \Bigg( \frac{h}{8 \pi \text{Kn}'} \Bigg) (B_2 + i k_2 A z ) e^{-kz},\tag|{56b} \]

\[ \hat{u}_3'' = \Bigg( \frac{h}{8 \pi \text{Kn}'} \Bigg) (B_3 +  k A z ) e^{-kz}. \tag|{56c}\]

\noindent
Adding the relevant components for the velocity and stress tensor fields due to a Stokeslet and a mirror image Stokeslet in the perpendicular direction, we obtain

\[  u_1{^{\perp}} + u_1{^{\perp}}' = - \frac{h}{4 \pi \text{Kn}'} \frac{x}{r_h^3}  , \tag|{57a}\]

\[ u_2{^{\perp}} + u_2{^{\perp}}' = - \frac{h}{4 \pi \text{Kn}'} \frac{y}{r_h^3} , \tag|{57b}\]

\[u_2{^{\perp}} + u_2{^{\perp}}' = S_{31}^{\perp} + S_{31}{^{\perp}}' = S_{32}{^{\perp}} + S_{32}{^{\perp}}' = 0.\tag|{57c}\]

\noindent
The Fourier transformed boundary conditions are

\[ \hat{u}_1{^{\perp}} + \hat{u}_1{^{\perp}}' + \hat{u}_1{^{\perp}}'' = - \sqrt{\frac{\pi}{2}} ( \hat{S}_{31}{^{\perp}} + \hat{S}_{31}{^{\perp}}' + \hat{S}_{31}{^{\perp}}''), \tag|{58a}\]

\[\hat{u}_2{^{\perp}} + \hat{u}_2{^{\perp}}' + \hat{u}_2{^{\perp}}'' = - \sqrt{\frac{\pi}{2}} ( \hat{S}_{32}{^{\perp}} + \hat{S}_{32}{^{\perp}}' + \hat{S}_{32}{^{\perp}}''), \tag|{58b}\]

\[ \hat{u}_3{^{\perp}} + \hat{u}_3{^{\perp}}' + \hat{u}_3{^{\perp}}'' = 0 . \tag|{58c}\]

\noindent
Taking Fourier transforms of (57) and adding everything at the boundary at $z=0$, we obtain

\[ \hat{u}_1{^{\perp}}'' - \text{Kn}\frac{\partial \hat{u}_1{^{\perp}}''}{\partial z}  = i \frac{h}{4 \pi \text{Kn}'}\frac{k_1}{k}, \tag{59a}\]

\[ \hat{u}_2{^{\perp}}'' - \text{Kn}\frac{\partial \hat{u}_2{^{\perp}}''}{\partial z}  = i \frac{h}{4 \pi \text{Kn}'}\frac{k_2}{k}, \tag{59b}\]

\[ \hat{u}_3{^{\perp}}'' = 0 . \tag{59c}\]

\noindent
Closing the system with the same Fourier transformed continuity equation as before, we obtain for the coefficients

\[  B_1^{\perp} = \frac{2 i k_1}{k(1 + 2  k \text{Kn})} ,  \tag|{60a} \]

\[ B_2^{\perp} = \frac{2 i k_2}{k(1 + 2  k \text{Kn})}  , \tag|{60b} \]

\[  B_3^{\perp} = 0 ,\tag|{60c} \]

\[  A^{\perp} = \frac{- 2}{1 + 2  k \text{Kn}}  .\tag|{60d}  \]

\noindent
Note that the Knudsen number is now the standard Knudsen number because the scaling has cancelled out.  After dimensionalisation to account for the slip length $\lambda$, these are the same Fourier coefficients as the ones obtained in [39] ie. when $\text{Kn} $ is replaced by $\lambda$.  In fact, the Fourier coefficients are the same as those in the no-slip case (when the slip length $\lambda $ is zero) divided through by $1 + 2 \lambda k$, so by differentiating one can write down the relation

\[ \Bigg( 1 - 2 \lambda \frac{\partial}{\partial z} \Bigg) \hat{p}{^{\perp}}'' (\textbf{r}, \lambda) = \hat{p}{^{\perp}}'' (\textbf{r},0). \tag|{61}\]

\noindent
This implies a differential equation for the inverted quantities:

\[ \Bigg( 1 - 2 \lambda \frac{\partial}{\partial z} \Bigg) p{^{\perp}}'' (\textbf{r}, \lambda) = p{^{\perp}}'' (\textbf{r},0). \tag|{62}\]

\noindent
In [38], it is shown that

\[ p{^{\perp}}''(\textbf{r},0) = \frac{h}{2 \pi} \frac{\partial}{\partial z} \Bigg(\frac{z}{|\textbf{r}|^3} \Bigg), \tag|{63}\]

\noindent
so integrating by parts and adding the other contributions $p^{\perp}$ and $p{^{\perp}}'$ due to the Stokeslet and the mirror image Stokeslet, we find that the total pressure due to a point forcing in Stokes flow close to and perpendicular to a flat slip boundary is

\[p^{\perp} = \frac{1}{4 \pi} \Bigg( \frac{z}{|\textbf{r}|^3} - \frac{z}{|\overline{\textbf{r}}|^3} + \frac{h}{\lambda} \int^{\infty}_0 \text{d} s \: \Bigg[ \frac{\partial}{\partial z} \Bigg(\frac{z}{|\textbf{r}|^3} \Bigg) \Bigg] (\textbf{r} + (h+s) \textbf{e}_z) e^{-s/2 \lambda}   \Bigg) ,\tag|{64}\]

\noindent
where $\textbf{e}_z$ is the unit vector in the $z$-direction.

\subsection{Temperature Distribution for a Thermal Stokeslet close to a Temperature Jump Surface}

\noindent
Finally, we consider the case of a point heat source (thermal Stokeslet) in heat flow close to a temperature jump surface (equations (14), solutions (17) and boundary condition (19)).  By the same analysis as before, one obtains for the Fourier transformed temperature

\[\hat{\theta} = A e^{-kz}. \tag|{65}\]

\noindent
The Fourier transformed boundary condition is

\[  \hat{\theta} + \hat{\theta}' + \hat{\theta}'' = - \frac{1}{2} \sqrt{\frac{\pi}{2}}( \hat{q}_3 + \hat{q}_3' + \hat{q}_3'' ) . \tag|{66}\]

\noindent
By the same or very similar analysis as Section 3.1, we substitute in the Fourier transformed quantities at the boundary where $z=0$ and then rearrange for the coefficient $A$, which is found to be

\[A = -\frac{2g}{15 \text{Kn}'\pi k (1 + \frac{15}{8} \text{Kn} k)} . \tag|{67}\]

\noindent
Note that this is the same Fourier coefficient as the one which we obtained for the same problem in Section 3.1 with a no-jump boundary after dividing through by $1 + 15\text{Kn}k/8$.  Differentiating as before implies a differential equation for $\hat{\theta}''$, which is inverted to give

\[ \Bigg( 1 - \frac{15}{8} \text{Kn} \frac{\partial}{\partial z} \Bigg) \theta'' = \theta_{nj}'', \tag|{68}\]

\noindent
where $\theta_{nj}''$ is the unknown temperature field for the heat source close to the no temperature jump surface.  Substituting in $\theta''_{nj}$ which was found in Section 3.1, integrating by parts and adding $\theta''$ to the temperature fields $\theta$ and $\theta'$ contributed by the heat source and the mirror image heat source, we obtain the total temperature field due to a point heat source placed in isobaric stationary Stokes flow close to a temperature jump boundary:

\[ \theta = \frac{g}{15 \text{Kn}' \pi} \Bigg( \frac{1}{|\textbf{r}|} -  \frac{1}{|\overline{\textbf{r}}|} + \frac{16}{15 \text{Kn}} \int^{\infty}_0 \text{d}s \: \Bigg[ \frac{1}{{|\textbf{r}|}} \Bigg] (\textbf{r} + (h+s) \textbf{e}_z ) e^{-8 s/ 15 \text{Kn}}   \Bigg). \tag|{69}\]

\noindent
The components of the heat flux can be obtained in a similar way.

\section{Solutions for Rarefied Gas Flows close to Boundaries}

\noindent
In this section, we will return to rarefied gas flows and heat flows modelled by the Grad equations.  We will find in this section that the method of images can be pushed far enough to obtain the solution in the case of a point heat source in a heat flow close to a flat temperature jump surface, but that obtaining the solution in the case of a point forcing in a rarefied gas flow close to a flat velocity slip surface is intractable (even if we assume for simplicity that the point forcing is perpendicular rather than parallel to the boundary). 

\subsection{Temperature Distribution for a Thermal Gradlet close to a Temperature Jump Surface}

\noindent
We start with the case of a point heat source (thermal Gradlet) placed in a heat flow close to a temperature jump surface (equations (26), solutions (28) and boundary condition (30)).  The Fourier transformed temperature jump boundary condition is now quite complicated:  

\[ \hat{\theta} + \hat{\theta}' + \hat{\theta}'' =-\frac{1}{2} \sqrt{\frac{\pi}{2}} ( \hat{q}_3 + \hat{q}_3' + \hat{q}_3'' ) - \frac{1}{4} (\hat{S}_{33} + \hat{S}_{33}' + \hat{S}_{33}'').\tag|{70} \]

\noindent
If we Fourier transform the equations (26) and solve the resulting differential equations, we obtain

\[\hat{\theta}'' = A e^{-kz} , \tag|{71a}\]

\[\hat{q}_3'' =  \frac{15}{4}k \text{Kn}' A e^{-kz}, \tag|{71b}\]

\[ \hat{S}_{33}'' = 3 \text{Kn}'{^2} k^2 A e^{-kz} , \tag|{71c}\]

\noindent
Using the solutions (28), we also have at the boundary

\[ \theta + \theta' = \frac{2 g}{15 \text{Kn}' \pi r_h} ,\tag|{72a}\]

\[ q_3 + q_3' =0, \tag|{72b}\]

\[S_{33} + S_{33}' = \frac{2 \text{Kn}' g}{5 \pi} \Bigg( \frac{3h^2}{r_h^5} - \frac{1}{r_h^3} \Bigg). \tag|{72c}\]

\noindent
Taking Fourier transforms of the above and adding everything at the boundary, we arrive after some rearrangements at

\[ A = - \frac{2g + \frac{15}{10}\text{Kn}'{^2} g k^2}{15 \text{Kn}' \pi k ( 1 + \frac{15}{8}\text{Kn}k + \frac{3}{4} \text{Kn}'{^2} k^2)}. \tag|{73}\]

\noindent
Substituting this back into (71a) and differentiating, we obtain after some algebra the following relation 

\[ \Bigg( 1 - \frac{15}{8} \text{Kn} \frac{\partial}{\partial z} + \frac{3}{4} \text{Kn}'{^2} \frac{\partial^2}{\partial z^2} \Bigg) \hat{\theta}'' = - \frac{2g}{15 \text{Kn}' \pi k}e^{-kz} - \frac{ \text{Kn}' g k}{10 \pi} e^{-kz} .\tag|{74} \]

\noindent
This implies a differential equation for the unknown temperature field $\theta''$ after inverting:

\[ \Bigg( 1 - \frac{15}{8} \text{Kn} \frac{\partial}{\partial z} + \frac{3}{4} \text{Kn}'{^2} \frac{\partial^2}{\partial z^2} \Bigg) \theta'' = -\frac{\text{Kn}' g}{10 \pi} \Bigg( \frac{1}{|\textbf{r}|^3} \Bigg( \frac{3 z^2}{|\textbf{r}|^2} - 1 \Bigg) \Bigg)    -\frac{2g}{15 \text{Kn}' \pi} \frac{1}{|\textbf{r}|}. \tag|{75}\]

\noindent
Solving this linear PDE for $\theta''$, we obtain an explicit solution

\[ \theta'' = \frac{16 g}{15 C \text{Kn}'{^2} \pi} \Bigg(  \int_0^{\infty} \text{d}s   \: K e^{-(15 \sqrt{2} \sqrt{1/\pi} + C)s/24 \text{Kn}' } (\textbf{r} + (h+s) \textbf{e}_z) - \int_0^{\infty} \text{d}s   \: K e^{(-15 \sqrt{2} \sqrt{1/\pi} + C)s/24 \text{Kn}' } (\textbf{r} + (h+s) \textbf{e}_z)  , \tag{76} \]

\noindent
where
\[C = \sqrt{450 \pi - 768}, \tag|{77a} \]

\[K = \frac{  (3 \text{Kn}'{^2} - 4y^2 -4z^2)x^2 + (3 \text{Kn}'{^2} - 4z^2) y^2 - 6 \text{Kn}'{^2} z^2 - 2x^4 - 2y^4 - 2z^4   }{|\textbf{r}|^5}. \tag|{77b}\]

\noindent
The two arbitrary constants which appear in this solution vanish, because the temperature goes to zero at infinity and the constants both multiply exponential terms raised to a positive multiple of $z$.  Adding $\theta''$ to $\theta $ and $\theta'$ finally results in the total temperature field due to a point heat source which is placed close to a flat temperature jump surface.

\subsection{Gradlet perpendicular to a Slip Surface}

\noindent
We will end our discussion of the method of images by considering the case of a point forcing placed in a rarefied gas flow close to and perpendicular to a slip surface (equations (21) and (22), solutions (23) and boundary conditions (25)).  $\hat{p}{^{\perp}}''$, $\hat{u}_1{^{\perp}}''$, $\hat{u}_2{^{\perp}}''$ and $\hat{u}_3{^{\perp}}''$ are the same as in Section 3.2 where we studied the point forcing close to a no-slip boundary.  The Fourier transformed boundary conditions are

\[u_1{^{\perp}} + u_1{^{\perp}}' + u_1{^{\perp}}'' =  - \sqrt{\frac{\pi}{2}}( S_{3 1}^{\perp} + S_{3 1}{^{\perp}}' + S_{3 1}{^{\perp}}''  ) - \frac{1}{5} (q_{1}^{\perp} + q_{1}{^{\perp}}' + q_{1}{^{\perp}}''     ), \tag|{78a}\]

\[u_2^{\perp} + u_2{^{\perp}}' + u_2{^{\perp}}'' = - \sqrt{\frac{\pi}{2}}( S_{3 2}^{\perp} + S_{3 2}{^{\perp}}' + S_{3 2}{^{\perp}}''  ) - \frac{1}{5} (q_{2}^{\perp} + q_{2}{^{\perp}}' + q_{2}{^{\perp}}'' )  , \tag|{78b}\]

\[u_3^{\perp} + u_3{^{\perp}}' +u_3{^{\perp}}''= 0 . \tag|{78c}\]

\noindent
Using equations (21), we obtain

\[  u_1 + u_1{^{\perp}}'  =  - \frac{1}{4 \pi \text{Kn}'} \frac{xh}{r_h^3} - \frac{9 \text{Kn}'}{10 \pi} \frac{xh}{r_h^5}, \tag|{79a}\] 

\[  u_2 + u_2{^{\perp}}'  =  - \frac{1}{4 \pi \text{Kn}'} \frac{yh}{r_h^3} - \frac{9 \text{Kn}'}{10 \pi} \frac{yh}{r_h^5}, \tag|{79b}\] 

\[ u_3 + u_3{^{\perp}}'  =0 , \tag|{79c}\] 

\[ q_1^{\perp} + q_1{^{\perp}}' = -\frac{9 \text{Kn}'}{8 \pi} \frac{hx}{r_h^5}  , \: \: \: \: q_2^{\perp} + q_2{^{\perp}}' = -\frac{9 \text{Kn}'}{8 \pi} \frac{hy}{r_h^5}, \tag|{79d} \]

\[  S_{31}^{\perp} + S_{31}{^{\perp}}'= S_{32}^{\perp} + S_{32}{^{\perp}}' =0  .\tag|{79d}\]

\noindent
Fourier transforming the equation (22a) for the stress tensor $\textbf{S}$ we get

\[ \hat{S}''_{ 3 1} =  - \text{Kn} \Bigg( \frac{\partial \hat{u}''_1}{\partial z} - i k_1 \hat{u}''_3 \Bigg)  - \frac{4}{5}\text{Kn}' \Bigg( \frac{\partial \hat{q}''_1}{\partial z} - i k_1 \hat{q}''_3 \Bigg)  \tag|{80a}  \]


\[ \hat{S}''_{ 3 2} = - \text{Kn} \Bigg( \frac{\partial \hat{u}''_2}{\partial z} - i k_2 \hat{u}''_3 \Bigg) - \frac{4}{5} \text{Kn}' \Bigg( \frac{\partial \hat{q}''_2}{\partial z} - i k_2 \hat{q}''_3 \Bigg),\tag|{80b} \]

\noindent
It is possible to Fourier transform equations (80) and add everything at the boundary as we have now done several times throughout this article, but after solving for the Fourier coefficients, the problem of taking inverse two-dimensional Fourier transforms is completely intractable. 

\section{Solution for Unsteady Thermal Gradlet in One Dimension}

\noindent
So far all the flows which we have studied have been time-independent.  In dimensions two and three, it is also possible to study the solution to the unsteady Stokes equations with a time-dependent point forcing or point heat source applied to the conservation equations, but the analysis is somewhat involved [19, 23].  The next question is whether one can solve the unsteady Grad equations with a time-dependent point heat source.  In order to simplify the equations far enough that they can be solved in this setting, we will have to assume one dimension: this is a standard assumption found elsewhere in the literature on the unsteady Grad and R13 equations [40].   We start with the linearised Grad equations from [28] in one dimension with a time-dependent point heat source applied to the conservation of energy equation.  This gives us the linearised dimensionless conservation equations   

\[ \frac{\partial \rho}{\partial t} + \frac{\partial v_1}{\partial x} = 0  , \tag|{81a}\] 

\[ \frac{\partial v_1}{\partial t} + \frac{\partial S_{11}}{\partial x} + \frac{\partial p}{\partial x} = 0 , \tag|{81b}\]

\[ \frac{3}{2} \frac{\partial \theta}{\partial t} + \frac{\partial v_1}{\partial x} + \frac{\partial q_1}{\partial x} = g \delta(x) \delta(t)    , \tag|{81c}\]

\noindent
along with the linearised dimensional equations for the heat flux and the stress tensor

\[ \frac{\partial S_{11}}{\partial t}  + \frac{8}{15} \text{Pr} \frac{\overline{w}_3}{\overline{w}_2} \frac{\partial q_1}{\partial x} = - \frac{2}{\overline{w}_2} \frac{1}{\text{Kn}}  \Bigg( S_{11} + \frac{4}{3} \text{Kn} \frac{\partial v_1}{\partial x}  \Bigg), \tag|{82a} \]

\[ \frac{\partial q_1}{\partial t} + \frac{5}{4 \text{Pr}} \frac{\theta_4}{\theta_2} \frac{\partial S_{11}}{\partial x} = - \frac{1}{\theta_2} \frac{5}{2 \text{Pr}} \frac{1}{\text{Kn}} \Bigg( q_1 + \frac{5}{2 \text{Pr}} \text{Kn} \frac{\partial \theta}{\partial x} \Bigg)  , \tag|{82b}\]

\noindent
where $\delta$ is the Dirac delta function and $g$ is the strength of the point heat source.  The other coefficients are defined in [28] and depend on the collision model which is being used. The Prandtl number is $\text{Pr} = \mu c_p/k$, where $\mu$ is the shear viscosity, $c_p$ is the isobaric specific heat, and $k$ is the thermal conductivity.  For simplicity, we will take the values for Maxwell molecules [28]:

\[ \theta_2 = \frac{45}{8}, \:\:\:\:\: \overline{w}_2 = 2, \:\:\:\:\: \overline{w}_3 = \theta_4 = 3, \:\:\:\:\: \text{Pr} = \frac{2}{3}  . \tag|{83}\]

\noindent
As usual for the point heat source, we search for solutions with constant and uniform pressure and velocity.  From the ideal gas law, this implies that $\rho = - \theta$, so the equations become after some re-arranging

\[ \frac{\partial q_1}{\partial x} = g \delta (x) \delta (t) , \tag|{84a}\]

\[ \frac{\partial S_{11}}{\partial t} + \frac{8}{15} \frac{\partial q_1}{\partial x} = - \frac{1}{\text{Kn}} S_{11} , \tag|{84b}\]

\[ \frac{\partial q_1}{\partial t} =  \frac{2}{3} \frac{1}{\text{Kn}} q_1 + \frac{5}{2} \frac{\partial \theta}{ \partial x}  . \tag|{84c}\]

\noindent
It is then straightforward to take separate Laplace transforms with respect to $x$ and $t$, solve the resulting linear system for the Laplace transformed quantities $\hat{\theta}$, $\hat{q}_1$ and $\hat{S}_{11}$ and finish by taking inverse Laplace transforms to arrive at

\[q_1 = g \delta(t) , \tag|{85a}\]

\[ S_{11} = -\frac{8g}{15}e^{-t/\text{Kn}} \delta(x) , \tag|{85b}\]

\[ \theta = gx \Bigg( \frac{2}{5} \frac{\partial}{\partial t} \delta(t) - \frac{4}{15} \frac{1}{\text{Kn}} \delta(t) \Bigg) . \tag|{85c}\]

\noindent
The use of Laplace transforms for both variables makes sense as we only have first-order derivatives for space and time.  In the case of Stokes flow, the preference for using the Laplace transform for the time domain and the Fourier transform for the spatial domain is slightly subtle, and not simply so that one can solve an ODE of at most second order [41].  This solution could be employed in a mesh-free numerical method to model an extremely simple time-dependent heat flow such as the one-dimensional heat flow within a rarefied vapour phase situated between two liquid bodies [28].  An example of such a numerical method (the method of fundamental solutions) is described in [27].  This method enables one to model a flow using fundamental solutions for the equations of motion.  The boundary integral method also uses fundamental solutions but unlike MFS, this method requires the use of numerical integrations which can become computationally expensive.  In formal terms, MFS avoids use of a computational mesh by approximating the solution via a series of radial basis functions [42, 43].  As far as we aware, using the solution we have just derived in this method would not be too challenging, as the method has already been applied in three dimensions.  We also attempted to solve the unsteady Grad equations in one dimension with a time-dependent point forcing applied to the momentum equation, but found obtaining non-trivial solutions to be intractable.

\section{Grad Analogue of the Rotlet}

\noindent
In the literature on singularities of viscous flow, the point forcing is only the simplest possible type of singularity [30].  One can also obtain a higher-order singularity of Stokes flow corresponding to rotational motion: in other words, one can study the response of the flow field to a point torque rather than a point forcing.  This singularity is discussed in detail by Batchelor in the context of bulk stress in a suspension of non-spherical particles [44] and it should be simple to generalise to the Grad equations.  If one introduces a Stokeslet in the flow, it is well-known that one obtains higher-order singularities like the Stokes dipole (also called the doublet in the literature) and the Stokes quadrupole by taking the gradient [30].  Below, we do the same with the Gradlet and obtain the fields corresponding to a Grad dipole:

\[ \textbf{v} = \Bigg( \frac{\textbf{D} - \textbf{D}^T - \text{Tr} \textbf{D}}{|\textbf{r}|^3} + \frac{3 \textbf{r} \textbf{r}}{|\textbf{r}|^5} \Bigg) \frac{\textbf{r}}{8 \pi \text{Kn}'} -  \Bigg( \frac{\textbf{D} - \textbf{D}^T - \text{Tr} \textbf{D}}{|\textbf{r}|^5} + \frac{5 \textbf{r} \textbf{r}}{|\textbf{r}|^7} \Bigg) \frac{9 \text{Kn}' \textbf{r}}{20 \pi }     , \tag{86a} \]


\[ p = \frac{1}{4 \pi} \Bigg(  \frac{3 \textbf{r} \cdot \textbf{D} \textbf{r}}{|\textbf{r}|^5}    - \frac{\textbf{e} \cdot \textbf{D} \textbf{e}}{|\textbf{r}|^3}             \Bigg)         , \tag{86b}\]



\[  \textbf{q} = \Bigg( \frac{\textbf{D} - \textbf{D}^T - \text{Tr} \textbf{D}}{|\textbf{r}|^5} + \frac{5 \textbf{r} \textbf{r}}{|\textbf{r}|^7} \Bigg) \frac{9 \text{Kn}' \textbf{r}}{8 \pi}  ,            \tag{86c}                 \]


\noindent
where $\textbf{D}$ is a rank two tensor of strengths and $\textbf{e}$ is a unit vector.  

By analogy with the singularities of Stokes flow, we could refer to the symmetric part of the fields as the Grad stresslet (corresponding to straining motion) and the anti-symmetric part as the Grad rotlet (corresponding to rotational motion).  The streamlines due to the Grad rotlet are circles around the line through the origin.  In Figure 2, we plot the streamlines for the Stokes and Grad rotlets for comparison when $\text{Kn} =1$.  The colourmap and colour scale are the same for both figures, so one can see that the gradient for the Grad rotlet is much larger.  Note that the Grad rotlet has two terms, the first of which will dominate as $\text{Kn}$ tends to zero, leading to the streamlines which are observed for the Stokes rotlet and described in [3].  The rotlet singularity could have applications in biomedical science because it corresponds to a point torque.  If utilized in the correct way it could potentially capture angular momentum of cylindrical or aspherical dust particles close to boundaries in dilute gas flows (relevant for studying inhalation of particulates and droplets) [30].  The next step in that endeavour would be to derive the velocity fields when a Grad rotlet is placed close to a flat no-slip boundary in a rarefied gas flow (as mentioned earlier, a no slip boundary condition might still be a good approximation for certain microscale rarefied gas flows).  It would be desirable for realistic modelling purposes to derive the velocity fields associated to a Grad rotlet close to a slip surface in a rarefied gas flow, but we have already shown that this is an intractable problem even for the simpler case of a point forcing perpendicular to a slip surface.

\begin{figure*}
\centering
\includegraphics[width=0.5\textwidth]{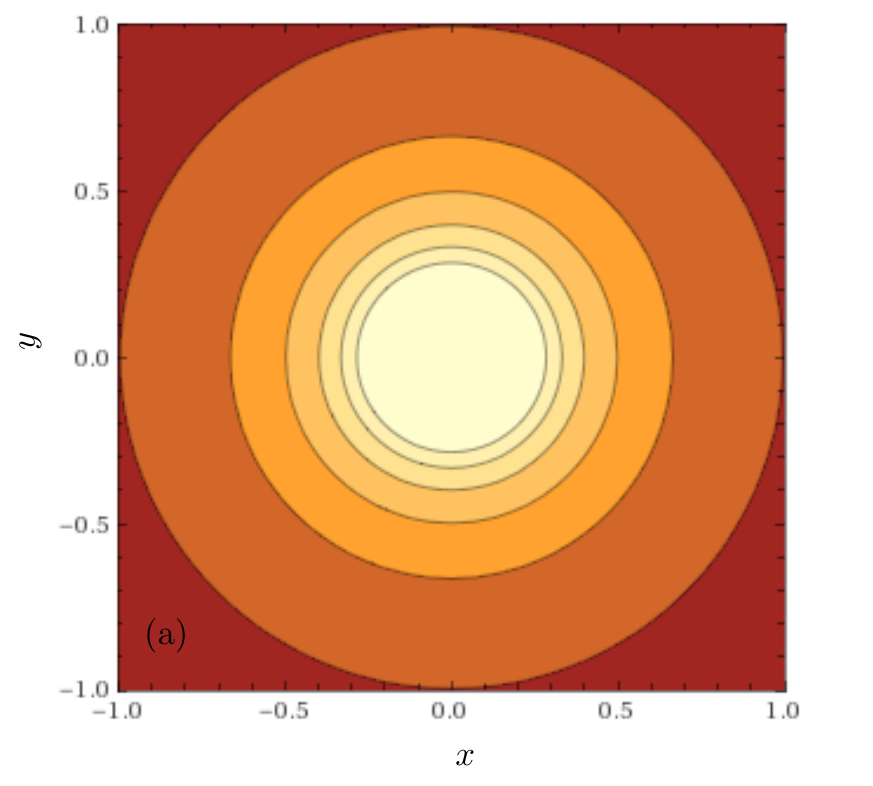} \includegraphics[width=0.5\textwidth]{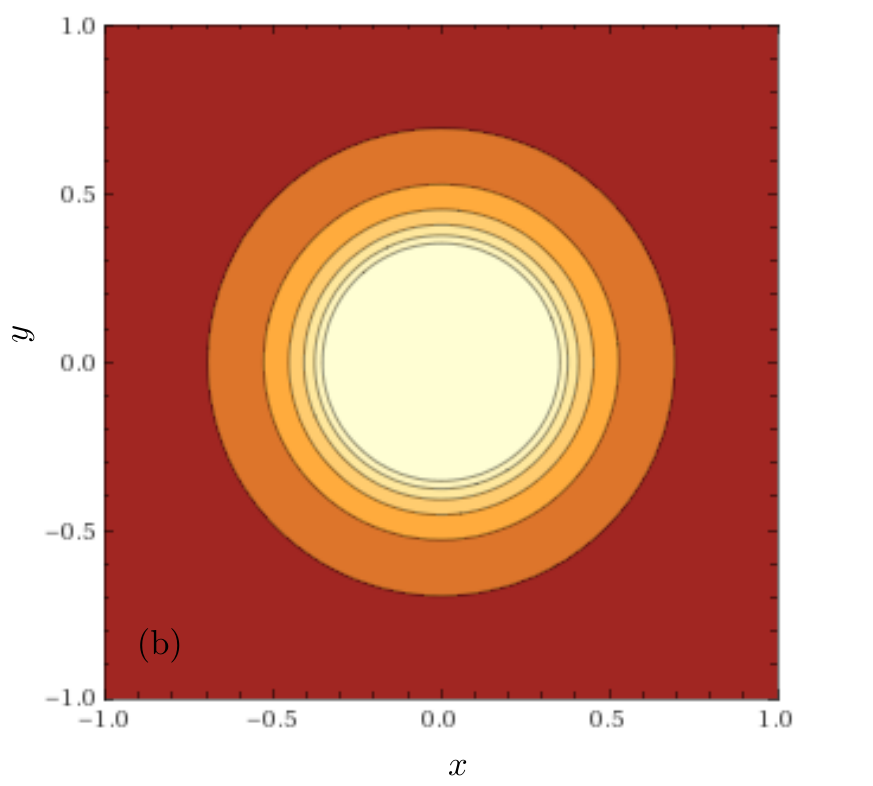}
\caption{Streamlines for (a.) Stokes Rotlet (b.) Grad Rotlet }
\end{figure*}

\section{Conclusions}

\noindent
In summary, we have studied some of the classical exact solutions and techniques from potential theory for Stokes flow (in particular, the method of images) and investigated the extent to which these carry over to rarefied microscale gas flows and heat flows modelled by the linearised Grad equations.  As a result we arrive at a number of new exact solutions which could be applied to the study of these flows in idealized geometries or used in numerical schemes for modelling purposes [37].  In particular, the integral solution for a thermal Gradlet close to a temperature jump surface could be employed in boundary integral techniques  for numerical solution of linear PDEs, where one computes flow and pressure fields by solving for distributions of singularities of the flow [25].  These numerical techniques are very familiar for Stokes flow calculations in physics and engineering and it might be interesting to use them for rarefied gas flow and heat flow calculations.  We have also derived several other solutions which could be employed in boundary integral techniques for rarefied gas flow computations (and in fact, the derivation of the pressure field even for the Stokeslet perpendicular to a slip surface is new as far as we know).  In some important respects, the techniques do not carry over completely and it is impossible to obtain an analogous solution for a point forcing placed in a rarefied gas flow close to a slip surface or a solution for the unsteady Grad equations in dimension higher than one.  As mentioned, the time-dependent solution to the unsteady Grad equations with a heat source could be employed to study very simple one dimensional heat flow problems [28] or it could be employed in a time-dependent mesh-free numerical method [27, 45].  It also seems possible that the Grad rotlet which we derived could find applications in biomedical science via the study of aspherical particles and droplets which move in dilute gas flows close to flat walls and have non-negligible angular momenta.

\section*{Acknowledgements} 
\noindent
The author would like to thank Alba Carballo Gonz{á}lez, Petr Denissenko and the two anonymous referees for their useful comments on the manuscript.  The author would also like to thank Jes{ú}s Fern{á}ndez Caballero for helpful discussions.  

\newpage
\section*{References}

\noindent
[1] Li H, Leong FY, Xu G, Ge Z, Kang CW and Lim KH (2020) Dispersion of evaporating cough droplets
in tropical outdoor environment.  Phys. Fluids 32, 113301

\noindent
[2] Cercignani C (2006) Slow Rarefied Flows: Theory and Application to Micro-electro-mechanical Systems.  Springer

\noindent
[3] Teolis BD, Jones GH, Miles PF, Tokar RL, Magee BA, Waite JH, Rousso E, Young DT, Crary FJ, Coates AJ, Johnson RE, Tseng W-L and Baragiola (2010) Cassini Finds an Oxygen-Carbon Dioxide Atmosphere at Saturn's Icy Moon Rhea.  Science 330, 6012, 1813-1815

\noindent
[4] Meng J, Zhang Y and Reese JM (2015) Numerical Simulation of Rarefied Gas Flows with Specified Heat Flux Boundary Conditions.  J. Fluid Mech. 17, 5, 1185-1200

\noindent
[5] Maxwell JC (1879) On stresses in rarefied gases arising from inequalities of temperature.  Phil. Trans. R. Soc. Lond A 170, 231-256

\noindent
[6] Bakanov S (1991) Thermophoresis in gases at small Knudsen numbers.  Aerosol Sci. Technol. 15(2), 77-92

\noindent
[7] Zheng F (2002) Thermophoresis of spherical and non-spherical particles: a review of theories and experiments.  Adv. Colloid Interface Sci. 97(1), 255-278

\noindent
\noindent
[8] Cercignani C and Daneri A (1963) Flow of a rarefied gas between two parallel plates.  J. Appl. Phys. 34(12), 3509-3513

\noindent
[9] Struchtrup H and Torrilhon M (2008) Higher-order effects in rarefied channel flows.  Physical Review E 8, 046301

\noindent
[10] Goswami P, Baier T, Tiwari S, Lv C, Hardt S and Klar A (2020) Drag force on spherical particle moving near a plane wall in highly rarefied gas.  J. Fluid Mech. 883:A47

\noindent
[11] Sone Y (2007) Molecular Gas Dynamics: Theory, Techniques, and Applications.  Birkhauser

\noindent
[12] Bird GA (1978) Monte Carlo simulation of gas flows.  Annu. Rev. Fluid. Mech, 10(1), 11-31.

\noindent
[13] Wu L, Reese J and Zhang Y (2014) Solving the Boltzmann equation deterministically by the fast spectral method: application to gas microflows.  J. Fluid Mech., 746:53-84.

\noindent
[14] Lockerby D and Collyer B (2016) Fundamental solutions to moment equations for the simulation of microscale gas flows.  J. Fluid Mech., 806:413-436

\noindent
[15] Sharipova F and Bertoldob G (2009) Poiseuille flow and thermal creep based on the Boltzmann equation with the Lennard-Jones potential over a wide range of the Knudsen number.  Phys. Fluids 21, 067101

\noindent
[16] Rosin MS, Caflisch RE, Dimits AM and Cohen BI (2011) Efficient Hybrid Methods for the Simulation of Plasmas with Coulomb Collisions.  American Physical Society, 53rd Annual Meeting of the APS Division of Plasma Physics.

\noindent
[17] Grad H (1949) On the kinetic theory of rarefied gases.  Commun. Pure Appl. Maths 2(4), 331-407

\noindent
[18] Chapman S and Cowling TC (1970) The Mathematical Theory of Non-Uniform Gases.  Cambridge University Press

\noindent
[19] Struchtrup H and Torrilhon M (2003) Regularization of Grad's 13 moment equations: Derivation and linear analysis.  Phys. Fluids 15, 9

\noindent
[20] Struchtrup H (2005) Macroscopic Transport Equations for Rarefied Gas Flows: Approximation Methods in Kinetic Theory.  Springer

\noindent
[21] Rana A, Torrilhon M and Struchtrup H (2013) A robust numerical method for the R13 equations of rarefied gas dynamics: Application to lid driven cavity.  J. Comput. Phys. 236, 169-186

\noindent
[22] Padrino JC, Sprittles JE and Lockerby DA (2019) Thermophoresis of a spherical particle: modelling through moment-based, macroscopic transport equations.  J. Fluid Mech. 862: 312-347

\noindent
[23] Young JB (2011) Thermophoresis of a spherical particle: reassessment, clarification and new analysis.  Aerosol Sci. Technol.  45(8), 927-948

\noindent
[24] Dwyer HA (1967) Thirteen-Moment Theory of the Thermal Force on a Spherical Particle.  Phys. Fluids 10: 976-984

\noindent
[25] Pozrikidis C (2010) Boundary Integral and Singularity Methods for Linearized Viscous Flow.  Cambridge University Press

\noindent
[26] Smith SH (1987) Unsteady Stokes flow in two dimensions.  J Eng. Math. 21:271-285

\noindent
[27]  Tsai CC, Young DL, Fan CM and Chen CW (2006) MFS with time-dependent fundamental solutions for unsteady Stokes equations.  Engineering Analysis with Boundary Elements 30(10), 897-908

\noindent
[28] Beckmann AF, Rana AS, Torrilhon M and Struchtrup H (2018) Evaporation Boundary Conditions for the Linear R13 equations based on the Onsager theory.  Entropy 20, 680

\noindent
[29] Zhicheng H, Siyao Y and Zhenning C (2020) Flows between parallel plates: Analytical solutions of regularized 13-moment equations for inverse-power-law models.  Phys. Fluids 32, 122007

\noindent
[30] Blake JR and Chwang AT (1974) Fundamental singularities of viscous flow.  J Eng, Math. 8:23-29

\noindent
[31] Sneddon IN (1951) Fourier Transforms.  McGraw-Hill

\noindent
[32] Erdelyi A (1954) A Table of Integral Transforms.  McGraw-Hill

\noindent
[33] Ladyzhenskaya OA (1963) The mathematical theory of viscous incompressible flow.  Gordon and Breach

\noindent
[34] Gu XJ and Emerson DR (2007) A computational strategy for the regularized 13 moment equatios with enhanced wall-boundary conditions.  J. Comput. Phys. 225(1):263-283

\noindent
[35] Lauga E, Brenner MP and Stone HA (2005) The no-slip boundary condition: a review, in Handbook of Experimental Fluid Dynamics, Springer

\noindent
[36] Muntz EP (1989) Rarefied gas dynamics.  Ann. Rev. Fluid Mech. 21:387-417

\noindent
[37] Claydon R, Srestha A, Rana AS, Sprittles JE and Lockerby DA (2017) Fundamental solutions to the regularised 13-moment equations: Efficient computation of three-dimensional kinetic effects.  J. Fluid Mech., 833:R4

\noindent
[38] Blake JR (1971) A note on the image system for a stokeslet in a no-slip boundary.  Proc. Camb. Phil. Soc., 70:303-310

\noindent
[39] Lauga E and Squires TM (2005) Brownian motion near a partial slip boundary: a local probe of the no-slip condition.  Phys. Fluids 17, 103102

\noindent
[40] Rana AS and Struchtrup H (2016) Thermodynamically admissible conditions for the regularized 13 moment equations.  Phys. Fluids 28 027105

\noindent
[41] Briggs KL (1964) Electron-stream Interaction with Plasmas.  MIT Press Research Monographs 29.  MIT Press 

\noindent
[42] Kupradze VD and Aleksidze MA (1964) The method of functional equations for the approximate solution of certain boundary value problems.  USSR Comput Math Math Phys 4:82-126

\noindent
[43] Mathon R and Johnston RL (1977) The approximate solution of elliptic boundary-value problems by fundamental solutions.  SIAM J Num Anal 14:638–50

\noindent
[44] Batchelor GK (1969) The stress system in a suspension of force-free particles.  J. Fluid Mech. 41(3):545-570

\noindent
[45] Williams H (2021) A Note on the Method of Fundamental Solutions.  Acad.  Lett. 745


\end{document}